\documentclass[aps,pra,reprint,superscriptaddress,twocolumn,longbibliography]{revtex4-2}
\usepackage{amsfonts,amssymb,amscd,amsthm,amsmath}
\usepackage{graphicx}
\usepackage{mathrsfs}
\usepackage[colorlinks, allcolors=blue]{hyperref}
\usepackage{etoolbox}
\usepackage{upgreek}

\setcitestyle{super}


\begin{document}

\title{Quantum simulation of a nonuniform Dicke model in a two-dimensional trapped-ion crystal}

\author{S.-A. Guo}
\thanks{These authors contribute equally to this work}%
\affiliation{Center for Quantum Information, Institute for Interdisciplinary Information Sciences, Tsinghua University, Beijing 100084, PR China}

\author{J. Ye}
\thanks{These authors contribute equally to this work}%
\affiliation{Center for Quantum Information, Institute for Interdisciplinary Information Sciences, Tsinghua University, Beijing 100084, PR China}

\author{J.-Y. Tan}
\affiliation{Center for Quantum Information, Institute for Interdisciplinary Information Sciences, Tsinghua University, Beijing 100084, PR China}

\author{Z.-W. Zhang}
\affiliation{Chizhi Institute, Shanghai, PR China}

\author{L. Zhang}
\affiliation{Center for Quantum Information, Institute for Interdisciplinary Information Sciences, Tsinghua University, Beijing 100084, PR China}

\author{Y.-Y. Chen}
\affiliation{Center for Quantum Information, Institute for Interdisciplinary Information Sciences, Tsinghua University, Beijing 100084, PR China}

\author{Y.-L. Xu}
\affiliation{Center for Quantum Information, Institute for Interdisciplinary Information Sciences, Tsinghua University, Beijing 100084, PR China}

\author{C. Zhang}
\affiliation{HYQ Co., Ltd., Beijing 100176, PR China}

\author{Y. Jiang}
\affiliation{Center for Quantum Information, Institute for Interdisciplinary Information Sciences, Tsinghua University, Beijing 100084, PR China}

\author{B.-X. Qi}
\affiliation{Center for Quantum Information, Institute for Interdisciplinary Information Sciences, Tsinghua University, Beijing 100084, PR China}

\author{L. He}
\affiliation{Center for Quantum Information, Institute for Interdisciplinary Information Sciences, Tsinghua University, Beijing 100084, PR China}
\affiliation{Hefei National Laboratory, Hefei 230088, PR China}

\author{Z.-C. Zhou}
\affiliation{Center for Quantum Information, Institute for Interdisciplinary Information Sciences, Tsinghua University, Beijing 100084, PR China}
\affiliation{Hefei National Laboratory, Hefei 230088, PR China}

\author{Y.-K. Wu}
\email{wyukai@mail.tsinghua.edu.cn}
\affiliation{Center for Quantum Information, Institute for Interdisciplinary Information Sciences, Tsinghua University, Beijing 100084, PR China}
\affiliation{Hefei National Laboratory, Hefei 230088, PR China}

\author{L.-M. Duan}
\email{lmduan@tsinghua.edu.cn}
\affiliation{Center for Quantum Information, Institute for Interdisciplinary Information Sciences, Tsinghua University, Beijing 100084, PR China}
\affiliation{Hefei National Laboratory, Hefei 230088, PR China}
\affiliation{New Cornerstone Science Laboratory, Institute for Interdisciplinary Information Sciences, Tsinghua University, Beijing 100084, PR China}

\begin{abstract}
Existing experimental realizations of the Dicke model in atomic or ionic systems are based on global observables assuming uniform spin-boson coupling, while inevitable experimental nonuniformity on the one hand requires site-resolved measurement of spin states, and on the other hand provides potential quantum advantage in simulating multi-spin distributions. Here we report the quantum simulation of a nonuniform Dicke-like model, due to either the inevitable experimental imperfection or the engineered spin-phonon coupling pattern, in a two-dimensional (2D) crystal of up to 200 ions. We explicitly demonstrate the sensitivity of few-spin observables and multi-spin distributions to the spatial inhomogeneity of the model, and examine the relaxation dynamics of the nonuniform model by measuring the subsystem entropies of selected ion groups. Our work enables the study of Dicke-like models beyond the symmetric subspace, paving the way toward understanding the role of nonuniformity in its thermalization and quantum chaos behavior.
\end{abstract}

\maketitle
\section*{Introduction}

The eigenstate thermalization hypothesis (ETH) provides a plausible explanation for how an isolated, nonintegrable quantum system thermalizes as in a microcanonical ensemble \cite{DAlessio03052016,Deutsch_2018,PhysRevA.43.2046,PhysRevE.50.888,Rigol2008}. Although the whole system undergoes unitary evolution with conserved population on each energy eigenstate, its small subsystems and thereby local observables generally approach the corresponding thermal equilibria rapidly and forget about the details of the initial states. ETH is deeply connected to quantum chaos \cite{DAlessio03052016,Deutsch_2018,PhysRevE.50.888,RevModPhys.83.863,PhysRevA.30.504} and quantum information scrambling \cite{Patrick_Hayden_2007,Yasuhiro_Sekino_2008,Swingle2018}, and has been widely studied theoretically \cite{PhysRevE.90.052105,PhysRevE.97.012140,PhysRevE.99.032111,PhysRevLett.128.180601} and tested experimentally in various physical systems \cite{doi:10.1126/science.1257026,doi:10.1126/science.aaa7432,doi:10.1126/science.aaf6725,Neill2016,Smith2016,PhysRevLett.117.170401,doi:10.1126/science.aau4963,doi:10.1126/science.abl6277,PRXQuantum.4.020318,PhysRevX.15.011035,doi:10.1126/science.ade7651,Andersen2025}, through both the thermalization dynamics of few-body observables \cite{doi:10.1126/science.1257026,doi:10.1126/science.aaa7432,doi:10.1126/science.aaf6725,Neill2016,Smith2016,PhysRevLett.117.170401,doi:10.1126/science.aau4963,doi:10.1126/science.abl6277,PRXQuantum.4.020318,PhysRevX.15.011035} and the global distribution of output states \cite{doi:10.1126/science.ade7651,Andersen2025}. Well-known exceptions to ETH such as many-body localization systems \cite{RevModPhys.91.021001,PhysRevLett.78.2803,BASKO20061126} and many-body scar states \cite{Serbyn2021,Moudgalya_2022} are also the subject of extensive research.

For spin systems, ETH is limited to subsystems less than half the size of the entire system by the maximal bipartite entanglement entropy \cite{DAlessio03052016,Deutsch_2018,PhysRevE.97.012140,PhysRevE.99.032111}. In more complex models such as the Dicke model \cite{PhysRev.93.99,HEPP1973360,PhysRevA.7.831,PhysRevA.8.1440,CARMICHAEL197347,PhysRevA.9.418,Kirton2019Introduction} where multiple spins are coupled to a bosonic mode, because of the infinite dimension of the bosonic system, the thermalization behavior may extend to a larger portion of the spins. Indeed, intricate thermalization and quantum chaos phenomena have been predicted in the Dicke model across different parameter regimes and for initial states with different energies \cite{Lewis-Swan2019,PhysRevResearch.3.L022020}, and have recently been demonstrated experimentally \cite{bullock2026quantum}. As a fundamental model in quantum optics describing light-matter interaction, the Dicke model has been realized experimentally in atomic \cite{Baumann2010,PhysRevLett.104.130401,PhysRevLett.107.140402,doi:10.1073/pnas.1417132112,Zhiqiang:17,PhysRevLett.113.020408} and ionic \cite{Cohn_2018,PhysRevLett.121.040503,bullock2026quantum} systems. However, existing studies typically consider only global observables, since resolving individual spins is experimentally challenging in a Bose-Einstein condensate or a cold atomic ensemble in an optical cavity, or in a fast-rotating ion crystal in a Penning trap. (Individual ions can be distinguished in a Penning trap for number counting \cite{britton2012engineered,bohnet2016quantum,bullock2026quantum}, but measurement of local spin observables has not yet been demonstrated.) Such global observables are sufficient to fully characterize the quantum state of spins, if we assume uniform spin-boson coupling in the Dicke model which ensures that the spin dynamics remains in the symmetric subspace known as the Dicke states. Nevertheless, as the system size increases, the inhomogeneity also becomes more significant, which may, for example, come from the spatial variations in the laser intensity or the trapping potential of trapped ions. Spatial inhomogeneity can further be engineered in systems like trapped ions by exploiting the rich structure of the collective phonon modes. On the other hand, going beyond the symmetric subspace, where the quantum dynamics can be simulated efficiently by a classical computer, also allows potential advantage for quantum simulation. These motivate us to investigate dynamics of nonuniform Dicke-like models using spatially resolved observables.

\begin{figure*}[!tbp]
	\centering
	\includegraphics[width=\linewidth]{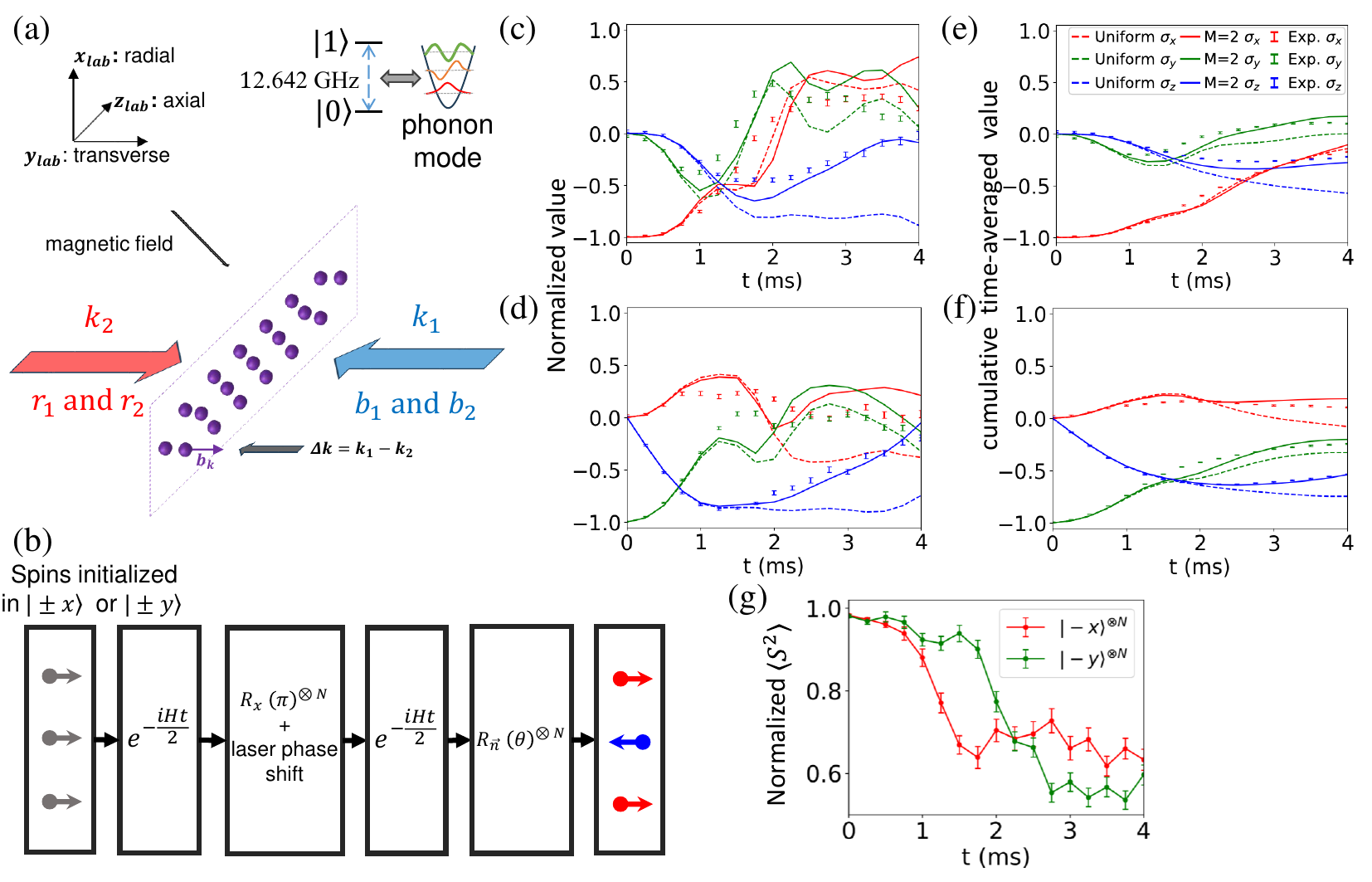}
	\caption {Experimental scheme. (a) Schematic of the experimental setup. We use counter-propagating $411\,$nm laser beams along the transverse ($y_{\mathrm{lab}}$) direction to couple the spin states of a two-dimensional crystal of up to $N=200$ ${}^{171}\mathrm{Yb}^+$ ions, located on the $x_{\mathrm{lab}}$-$z_{\mathrm{lab}}$ plane, to a shared collective phonon mode $k$. The wave vector difference $\boldsymbol{\Delta k}$ of the two laser beams and the mode vector $\boldsymbol{b_k}$ of the phonon mode are both in the $y_{\mathrm{lab}}$ direction. Each $411\,$nm laser beam contains two frequency components (red detunings $r_1$ and $r_2$ in one beam, and blue detunings $b_1$ and $b_2$ in the other beam) to generate spin-phonon coupling while compensating their time-independent AC Stark shift \cite{guo2024siteresolved}. A magnetic field is applied in the $y_{\mathrm{lab}}$-$z_{\mathrm{lab}}$ plane at $45^\circ$ to the crystal plane to maximize the spin-phonon coupling strength. (b) Experimental sequence. We initialize all the spins along the $x$ or $y$ direction, and then turn on the Dicke-like Hamiltonian $H_0$ for a controllable time $t$ with an SK1 composite pulse \cite{PhysRevA.70.052318} spin echo inserted in the middle. Finally, we use another SK1 composite pulse to rotate the spins globally into any desired basis with a controllable rotation axis $\vec{n}$ and rotation angle $\theta$, and measure the spins by electron shelving. (c-f) We verify the simulated Dicke-like model by dynamics of average single-spin observables $\sigma_x$ (red), $\sigma_y$ (green) and $\sigma_z$ (blue) from the initial spin states of $N=60$ ions in $|-x\rangle^{\otimes N}$ (top row) or $|-y\rangle^{\otimes N}$ (bottom row). The dashed lines are numerical simulation results for a homogeneous model, and the solid lines are computed by dividing the spins into two subensembles to account for the nonuniformity of the spin-phonon coupling. In the left column we present the raw dynamics which is sensitive to the shot-to-shot fluctuation of the experimental parameters, while in the right column we plot the cumulative time-averaged dynamics and get better agreement between experimental and theoretical results. (g) Evolution of the total spin $\langle S^2\rangle$, normalized by the maximal value $(N/2)(N/2+1)$ for the $N=60$ crystal, from the two initial states. Error bars represent one standard deviation from 200 experimental repetitions for each time point.} \label{fig1}
\end{figure*}

Here we report the quantum simulation of a nonuniform Dicke-like model using a two-dimensional (2D) ion crystal in a Paul trap with site-resolved readout \cite{guo2024siteresolved}. We start from small system sizes and few-body observables to verify the successful simulation of the model and the significance of the nonuniformity, and then extend to larger system sizes and many-body distributions that are challenging for direct classical simulation. We further extract subsystem entropies to characterize the relaxation dynamics of the nonuniform Dicke-like model, and quantify the positive correlation between the entropy-growth rates and the engineered spin-phonon coupling strengths. The observed relaxation dynamics are consistent with theoretical predictions, which indicate various thermalization behaviors for a homogeneous model in different parameter regimes and for different initial states. Our work provides a new paradigm for studying Dicke-like models beyond the idealized uniform-coupling limit, and can help understand the role of nonuniformity in their thermalization dynamics.

\section*{Results}
\subsection*{Nonuniform Dicke-like model}
Our experimental setup is sketched in Fig.~\ref{fig1}(a). We trap a two-dimensional (2D) crystal of ${}^{171}\mathrm{Yb}^+$ ions in a monolithic Paul trap at the temperature of $T=6\,$K \cite{guo2024siteresolved}. The spins are encoded in the $S_{1/2}$ hyperfine clock states as $|0\rangle\equiv |S_{1/2},F=0,m_F=0\rangle$ and $|1\rangle\equiv |S_{1/2},F=1,m_F=0\rangle$, and are coupled to a collective phonon mode of the ions by $411\,$nm laser beams perpendicular to the 2D crystal. Specifically, we use counter-propagating laser beams with a relative detuning $\mu$ to generate a spatial modulation of the AC Stark shift on the $|0\rangle$ state, thereby creating a spin-dependent force \cite{PhysRevA.103.012603}. We further apply symmetric frequency components on the two sides of the $|0\rangle \leftrightarrow |D_{5/2},F=2,m_F=0\rangle$ transition to compensate the time-averaged light shift \cite{guo2024siteresolved}. By setting the detuning $\mu$ close to the frequency $\omega_k$ of one of the collective phonon modes, we obtain the Hamiltonian for $N$ spins and one bosonic mode as \cite{guo2024siteresolved}
\begin{equation}
\begin{aligned}
H = & \eta \sum_{i=1}^N b_{i} \Omega_i (I+\sigma_z^i)(a e^{i\varphi_i}+a^\dag e^{-i\varphi_i}) \\
& - \delta a^\dag a + B\sum_{i=1}^N \sigma_x^i,
\end{aligned}
\label{eq:Hamiltonian}
\end{equation}
where $a$ and $a^\dag$ are the annihilation and creation operators for the chosen phonon mode with the Lamb-Dicke parameter $\eta$ and the laser detuning $\delta\equiv \mu-\omega_k$, and $\sigma_z^i\equiv |0\rangle_i\langle 0| - |1\rangle_i\langle 1|$ and $\sigma_x^i\equiv |0\rangle_i\langle 1| + |1\rangle_i\langle 0|$ are Pauli operators for the spin $i$. The spin-phonon coupling is governed by the site-dependent light shift $\Omega_i$ and the mode vector $b_i$ of the phonon mode, together with a spatially varying motional phase $\varphi_i$ which arises from the curvature of the 2D crystal with respect to the wavefront of the laser (see Methods for more details). In the following, we define $g_i\equiv \eta\sqrt{N}b_i\Omega_i$ to quantify the magnitude of the spin-phonon coupling, with a rescaling factor $\sqrt{N}$ to match the theoretical treatment in Eq.~(\ref{eq:subensemble}) in Methods. In addition, we apply a global microwave field resonant with the spin transition, which corresponds to a nearly uniform transverse field $B$. Note that, for $\delta>0$ ($\delta<0$), the Hamiltonian has an energy spectrum extending infinitely downward (upward). Because the thermalization dynamics typically depends on the location of the average energy within the spectrum, in the following, when $\delta<0$, we perform the transformation $H\to -H^*$, $|\pm x(z)\rangle\to|\pm x(z)\rangle$ and $|\pm y\rangle\to|\mp y\rangle$ which effectively obtains a positive $\delta$ without changing the dynamics. Note that here $|\pm x(y,z)\rangle$ are eigenstates of Pauli operators $\sigma_{x(y,z)}$ whose phase reference is set by the microwave drive. This should not be confused with the spatial directions $x_{\mathrm{lab}}$, $y_{\mathrm{lab}}$ and
$z_{\mathrm{lab}}$ in Fig.~\ref{fig1}(a).

Ideally, with a broad laser beam $\Omega_i=\Omega$, a uniform trapping potential in which the center-of-mass (COM) mode satisfies $b_i=1/\sqrt{N}$, and a 2D crystal aligned with the equiphase surface of the laser $\varphi_i=0$, the Hamiltonian in Eq.~(\ref{eq:Hamiltonian}) resembles the Dicke model apart from a spin-independent force. We can further perform a spin echo with a phase shift $\Delta\varphi_i=\pi$ of the laser, such that the spin-independent force gets cancelled for a short evolution time $t$ together with small uncompensated time-averaged light shift proportional to $\sigma_z^i$. We thus execute the experimental sequence in Fig.~\ref{fig1}(b) with the initial phonon state cooled close to the ground state by Doppler cooling, electromagnetically-induced-transparency cooling and sideband cooling \cite{guo2024siteresolved}, and the initial spin state $|\pm x\rangle^{\otimes N}$ or $|\pm y\rangle^{\otimes N}$ prepared by a global microwave SK1 composite $\pi/2$ pulse \cite{PhysRevA.70.052318} which compensates the nonuniformity of the microwave field over the large 2D ion crystal. For such a symmetric state under the exchange of spin labels, the subsequent evolution should ideally remain in the symmetric subspace. Then after a controllable evolution time $t$, we further perform a global rotation using another SK1 composite pulse, and measure the spins in different bases by electron shelving \cite{Roman2020,edmunds2020scalable}, from which we analyze their few-body observables or global distributions.
For example, to verify the successful realization of the Dicke-like model Hamiltonian, we set the laser detuning to be $\delta=2\pi\times 0.5\,$kHz above the COM mode and the transverse field $B=2\pi\times 0.09\,$kHz, and compare the measured average $\langle\sigma_x\rangle$, $\langle\sigma_y\rangle$ or $\langle\sigma_z\rangle$ over $N=60$ spins with the theoretical predictions as shown in Figs.~\ref{fig1}(c) and (d) for the initial spin states $|-x\rangle^{\otimes N}$ and $|-y\rangle^{\otimes N}$, respectively. As we can see, while the ideal homogeneous model (dashed curves) captures the rough tendency of these single-spin observables, we can largely improve the theoretical predictions by dividing the spins into two subensembles (solid curves) with different coupling strengths $\bar{g}_1 = 2\pi \times 0.23\,$kHz and $\bar{g}_2 = 2\pi \times 0.09\,$kHz, and different motional phases $\bar{\varphi}_1 = 0.056\pi$ and $\bar{\varphi}_2 = -0.056\pi$, to better represent the spatial nonuniformity of the spin-boson coupling (see Methods for more details). The remaining discrepancies are mainly attributed to the shot-to-shot fluctuation of experimental parameters such as the laser and microwave intensities and the trap frequency. In Figs.~\ref{fig1}(e) and (f) we present cumulative time-averaged observables $\overline{\sigma}(t)\equiv (1/t)\int_0^t \sigma(t^\prime)dt^\prime$ to suppress the sensitivity to the parameter drift and obtain better agreement between theoretical and experimental results without needing to model it as a decoherence source. Finally, in Fig.~\ref{fig1}(g) we show the evolution of the total spin $\langle S^2\rangle$, whose deviation from the maximal value $(N/2)(N/2+1)$ again quantifies the leakage outside the symmetric subspace due to the spatial nonuniformity in the Dicke-like model. As we can see, considerable leakage has been observed after $1$-$2\,$ms evolution.

\begin{figure*}[!tbp]
	\centering
	\includegraphics[width=\linewidth]{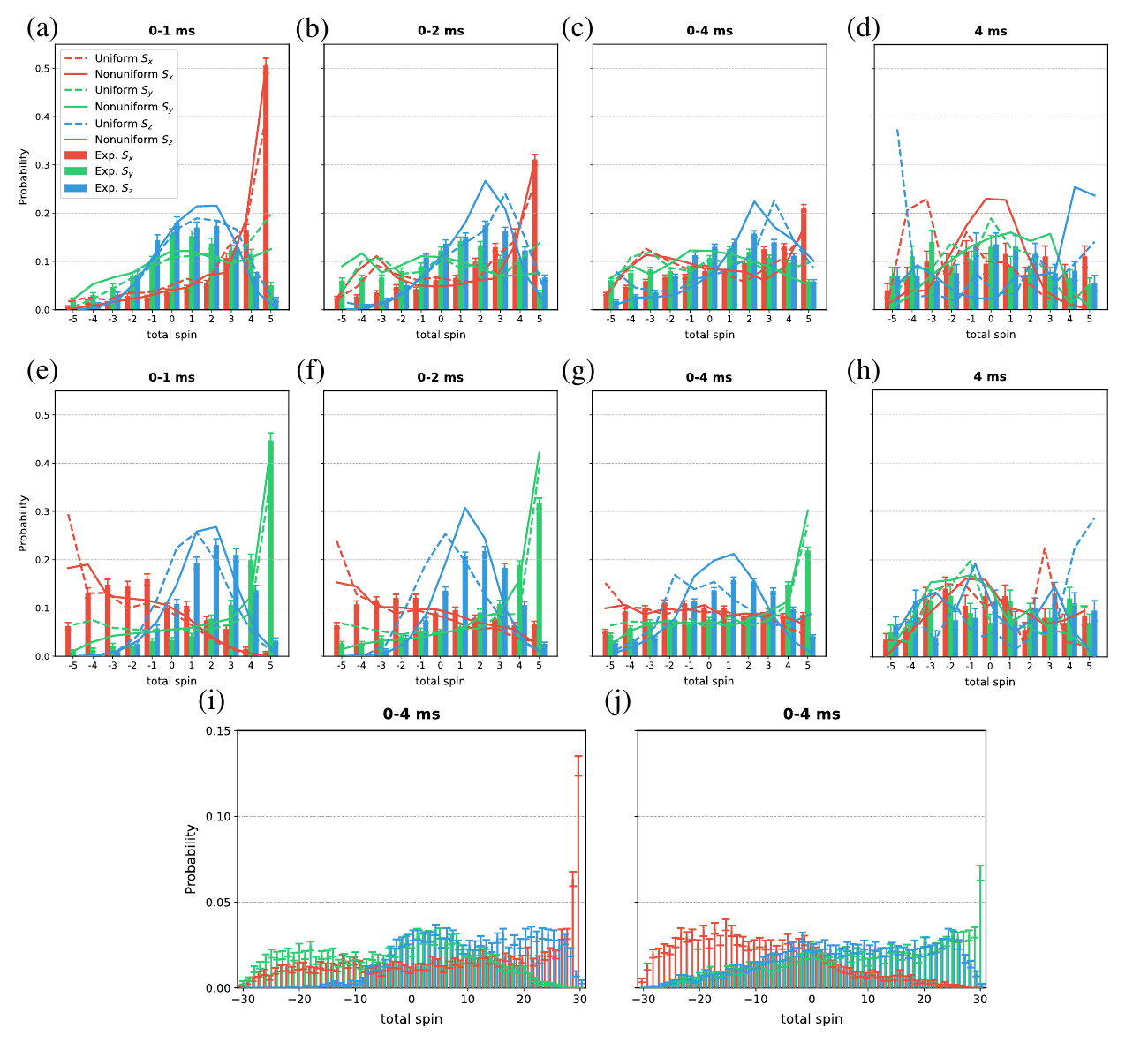}
	\caption{Relaxation dynamics of a Dicke-like model in global distributions. We couple the spins dominantly to the center-of-mass phonon mode, and measure the distributions of total spin components $S_x$ (red), $S_y$ (green) and $S_z$ (blue). (a)-(c) We initialize $N=10$ ions in $|+x\rangle^{\otimes N}$, and measure the cumulative time-averaged distributions over the intervals $0\le t\le T$ with $T=1,\,2,\,4\,$ms. The bars represent the experimental data, the dashed lines are numerical simulation results for an ideal homogeneous model, and the solid lines further consider the nonuniformity of the spin-phonon coupling. (d) Comparison between experimental and theoretical distributions at a single time point $t=4\,$ms. (e)-(h) Similar plots for $N=10$ ions initialized in $|+y\rangle^{\otimes N}$. (i) and (j) Similar cumulative time-averaged experimental results for $N=60$ ions initialized in $|+x\rangle^{\otimes N}$ and $|+y\rangle^{\otimes N}$, respectively. We take 200 samples for each time point and each measurement basis, and the error bars of the experimental data represent one standard error assuming a binomial distribution for each bin.
}\label{fig2}
\end{figure*}

\subsection*{Global spin distributions}
The nonnegligible effect of the inhomogeneity in the spin-boson coupling is also visible from global distributions of the measured spin states. In Fig.~\ref{fig2} we plot the histograms of the $x$, $y$ and $z$ components of the total spin $S_{x(y,z)}\equiv\sum_i \sigma_{x(y,z)}^i/2$. Note that in the presence of nonuniformity, the total spin is not a conserved quantity, but we can still use its components to group the output states as a coarse-grained way to visualize the global distributions. In the first and the second rows, we plot the experimentally measured distributions for $N=10$ ions evolved from the initial states $|+x\rangle^{\otimes N}$ and $|+y\rangle^{\otimes N}$, respectively, and compare with the theoretical distributions for an ideal Dicke model (dashed curves) and an exact spin-boson coupled model using the experimentally calibrated nonuniform laser intensities, motional phases and phonon mode structure (solid curves). For the homogeneous Dicke model in the weak-$B$ regime considered here ($B=2\pi\times 0.09\,$kHz), the spin part of the ground states are close to the ferromagnetic states $|\pm z\rangle^{\otimes N}$ \cite{Lewis-Swan2019}, whereas both $|+x\rangle^{\otimes N}$ and $|+y\rangle^{\otimes N}$ lie in the middle of the energy spectrum, so that we expect a similar relaxation behavior for them. In the first three columns, we take time points uniformly at $t=k\Delta t$ ($\Delta t=0.25\,$ms, $k=0,1,\cdots$) with 200 samples at each time point, and plot the cumulative time-averaged distributions. As the evolution time increases, the peak for the initial state flattens out. We emphasize that this flattening alone is not a proof of quantum thermalization, since cumulative time averaging can also wash out purely coherent oscillations. What we show is that the measured distributions are consistent with the theoretical calculations, which predict thermalization for the homogeneous model in this regime \cite{Lewis-Swan2019,PhysRevResearch.3.L022020}. The necessity of this cumulative time average for a comparison between the theoretical and the experimental results is due to the shot-to-shot parameter drift as mentioned above, which effectively already performs a time average on the experimental data. This can be seen more clearly at a single time point $t=4\,$ms as shown in Figs.~\ref{fig2}(d) and (h), where the measured distributions flatten even faster than the theoretical ones: The theoretical distributions still show residual peaks, which are however not visible in the experimental data. Furthermore, we observe that, even though here for the smaller $N=10$ ion crystal the spatial nonuniformity is also weaker, the exact model still performs slightly better than the uniform model, in particular for the initial state $|+y\rangle^{\otimes N}$. Also note that the global distributions considered here are generally more sensitive to experimental decoherence than the few-body observables in Fig.~\ref{fig1}, which explains the remaining deviation between the experimental and the theoretical results even after time averaging.

Finally, we also plot the measured distributions for $N=60$ spins in Figs.~\ref{fig2}(i) and (j) for initial states $|+x\rangle^{\otimes N}$ and $|+y\rangle^{\otimes N}$. The qualitative behavior is similar to that for $N=10$ ions, and the suppressed probabilities at high values of $|S_{x(y,z)}|$ can be explained by the increased spatial nonuniformity, which leads to the leakage from the original high-total-spin $S=N/2$ states to other low-total-spin states as indicated in Fig.~\ref{fig1}(g). In this case we do not present the theoretical results, because a homogeneous model or an approximate one with $M=2$ or $M=3$ subensembles will deviate significantly from the experimental results, while approximations at higher accuracy using more subensembles requires too high a Hilbert space dimension and is prohibitively expensive (see Methods for more details).

\begin{figure}[!tbp]
	\centering
	\includegraphics[width=\linewidth]{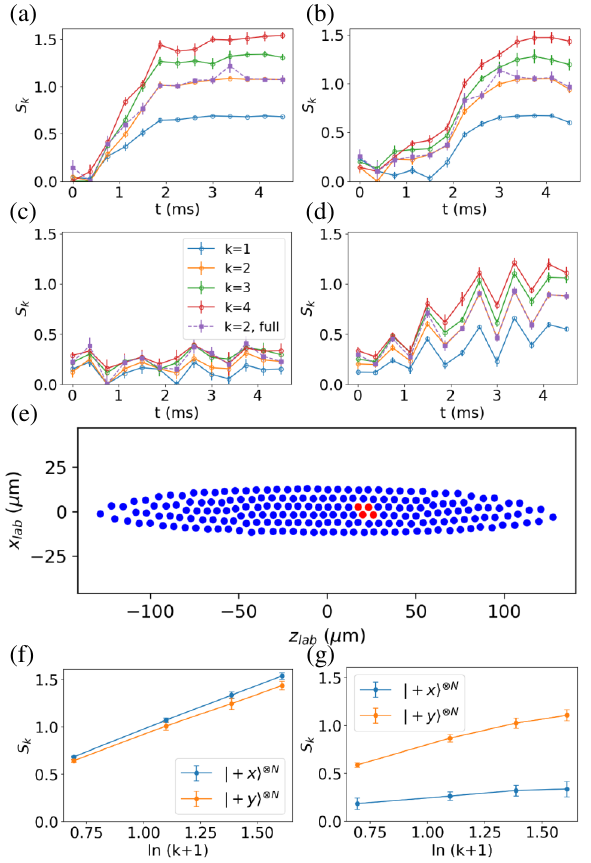}
	\caption{Evolution of subsystem entropy. We couple $N=200$ spins dominantly to the center-of-mass phonon mode, and measure the dynamics of the subsystem entropy under (a, b) a weak or (c, d) a strong transverse field, starting from the initial spin state (a, c) $|+x\rangle^{\otimes N}$ or (b, d) $|+y\rangle^{\otimes N}$. Here we choose a typical group of four neighboring ions as shown in (e) and measure the subsystem entropy $S_k$ for the $k=1,\,2,\,3,\,4$ spins assuming symmetric states as the circles connected by solid lines. We also compute the $k=2$ subsystem entropy by reconstructing the complete density matrices as the squares connected by the dashed lines. (f) Subsystem entropy $S_k$ versus subsystem size $k$ at the evolution time $t=4.5\,$ms in the weak-$B$ regime. (g) Similar plot in the strong-$B$ regime. Error bars represent one standard deviation estimated by bootstrap resampling.
} \label{fig3}
\end{figure}

\subsection*{Subsystem entropies and relaxation dynamics}
From the measured samples in various bases, we can further compute the subsystem entropy of arbitrarily selected groups of ions to characterize the relaxation dynamics of the Dicke-like model. For an ideal Dicke model, the $k$-spin states are restricted to a symmetric subspace with a dimension of $k+1$, hence we need about $O((k+1)^2)$ measurement bases to fully determine the reduced density matrix. This no longer holds for general nonuniform models, but for a small group of adjacent spins, we can expect vanishing spatial nonuniformity. Therefore the assumption of a symmetric subspace will still be valid locally, and we can reconstruct the reduced density matrices $\rho_k$ efficiently by the maximum likelihood method (see Methods for more details) and further compute the von Neumann entropy $S_k=-\mathrm{Tr}(\rho_k\ln\rho_k)$, where $\ln$ denotes the natural logarithm. In Figs.~\ref{fig3}(a)-(d), we present the subsystem entropies for $k=1,2,3,4$ adjacent ions in an $N=200$ crystal from the initial state $|+x\rangle^{\otimes N}$ (left column) or $|+y\rangle^{\otimes N}$ (right column), under a weak transverse field $B=2\pi\times 0.09\,$kHz (upper row) or a strong transverse field $B=2\pi\times 0.85\,$kHz (lower row) and a detuning $\delta=2\pi\times 0.5\,$kHz below the COM mode. To verify the assumption of the symmetric subspace, we also reconstruct the complete two-spin reduced density matrices for the $k=2$ cases by the maximum likelihood method using the measured data in random bases. The results are shown as the dashed curves and agree well with the above results. We also compute the standard deviation of the complex coupling strength $g_i e^{i\varphi_i}$ for the selected groups of two, three and four adjacent ions, and obtain $2\pi\times 8.3\,$Hz, $2\pi\times 7.0\,$Hz and $2\pi\times 8.3\,$Hz, respectively, which are more than an order of magnitude smaller than the standard deviation of $2\pi\times 134\,$Hz for the whole crystal. Further validation of the local symmetric-subspace approximation can be found in Methods.

Physically, the weak-$B$ regime corresponds to the superradiant phase of the Dicke model, and we observe a quick increase in the subsystem entropy up to some saturation time around $2$-$3\,$ms, consistent with the expected quantum chaos and thermalization behavior in a homogeneous model \cite{Lewis-Swan2019,PhysRevResearch.3.L022020}. In particular, given that in this parameter regime the Hamiltonian is dominated by the spin-phonon coupling in the $z$ basis, both initial states $|+x\rangle^{\otimes N}$ (with an average energy of $NB$) and $|+y\rangle^{\otimes N}$ (with an average energy of zero) lie in the middle part of the energy spectrum, and hence we expect similar relaxation behavior for them. On the other hand, in the strong-$B$ regime, namely the normal phase, we observe relatively slow entropy growth, in particular for an initial state polarized along the transverse field. This is because here $|+x\rangle^{\otimes N}$ is close to the upper edge of the energy spectrum, dominated by the transverse field, whereas $|+y\rangle^{\otimes N}$ still lies in the middle. This explains the strongly suppressed entropy growth from $|+x\rangle^{\otimes N}$ and the larger entropy growth rate from $|+y\rangle^{\otimes N}$ in Figs.~\ref{fig3}(c) and (d), although the growth rate for $|+y\rangle^{\otimes N}$ in this regime is nevertheless slower than those in the weak-$B$ regime, consistent with the theoretical prediction that the normal phase is nonthermal for a homogeneous model \cite{Lewis-Swan2019}.
Quantitatively, we plot the scaling of the subsystem entropy $S_k$ versus the subsystem size $k$ in Figs.~\ref{fig3}(f) and (g) at the evolution time $t=4.5\,$ms. In the weak-$B$ regime, we fit $S_k=\alpha \ln (k+1)$ with $\alpha_x^{(w)}=0.93$ and $\alpha_y^{(w)}=0.86$ for the two initial states, nearly saturating the entropy in the symmetric subspace. In comparison, in the strong-$B$ regime, we fit much lower $\alpha_x^{(s)}=0.17$ and $\alpha_y^{(s)}=0.57$.

Incidentally, in Fig.~\ref{fig3}(d) we observe periodic oscillations in the subsystem entropies, which originate from the strong transverse field combined with the weak spin-phonon coupling in Eq.~(\ref{eq:Hamiltonian}). Without the spin-phonon interaction, the transverse field would simply rotate all the spins globally and would not change the subsystem entropy. On the other hand, in a frame rotating with the transverse field, the spin-phonon interaction acquires a time-dependent spin phase oscillating at the frequency $2B$, which can entangle the spins with the phonon mode and increase the subsystem entropy. After one period $\pi/B\approx 0.6\,$ms, the accumulation of the spin-phonon entanglement is largely cancelled under the lowest-order Magnus expansion, which explains the observed periodic oscillation. Deviation from the theoretical period may be due to the nonuniformity and the long-term drift of the microwave intensity.

\begin{figure}[!tbp]
	\centering
	\includegraphics[width=\linewidth]{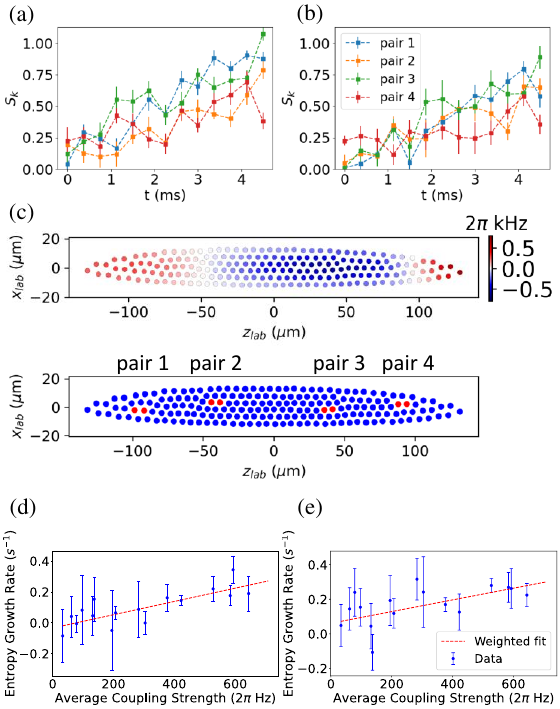}
	\caption{Evolution of subsystem entropy under engineered nonuniform spin-phonon coupling. We couple $N=200$ spins dominantly to the third highest phonon mode, and similarly measure the dynamics of the subsystem entropy starting from the initial spin state (a) $|+x\rangle^{\otimes N}$ or (b) $|+y\rangle^{\otimes N}$. We observe faster dynamics for ions with a stronger spin-phonon coupling (blue for pair 1 and green for pair 3) than for ions with a weaker spin-phonon coupling (orange for pair 2 and red for pair 4). (c) Experimentally calibrated spin-phonon coupling strength for individual ions and the locations of the ion pairs used in (a) and (b). (d) Fitted initial entropy-growth rates versus the effective coupling strengths $g$ for the initial state $|+x\rangle^{\otimes N}$. Here $g\equiv\sqrt{(g_1^2 + g_2^2)/2}$ is the root mean square of the spin-phonon coupling strengths $g_1$ and $g_2$ of an adjacent ion pair. The red dashed line is a linear fit, weighted by the inverse variance of each fitted entropy-growth rate. (e) Similar plot for the initial state $|+y\rangle^{\otimes N}$. Error bars represent one standard deviation. For (a) and (b) we use bootstrap resampling, and for (c) and (d) they are computed from ordinary least-squares regression.}
	\label{fig4}
\end{figure}

In the above examples, the ideal model is homogeneous and the nonuniformity arises from the experimental imperfections which typically vary slowly in space. However, the rich structure of the transverse phonon modes of the 2D ion crystal also allows us to purposely engineer a spin-boson coupled model with artificial spatial patterns. In Fig.~\ref{fig4}, we set the laser detuning to be $\delta=2\pi\times 0.75\,$kHz below the third highest phonon mode of an $N=200$ crystal, and similarly study the relaxation dynamics by subsystem entropy. Due to the engineered nonuniformity as shown in the upper panel of Fig.~\ref{fig4}(c), here we have a larger standard deviation of $2\pi\times 270\,$Hz of the complex coupling strength $g_i e^{i\varphi_i}$ over the whole crystal, about twice the value in Fig.~\ref{fig3}. Therefore, we expect the violation of the symmetric subspace assumption to be larger, and we choose to reconstruct the complete two-spin density matrices to evaluate the subsystem entropies. We choose four ion pairs at the locations indicated by the adjacent red dots in the lower panel of Fig.~\ref{fig4}(c), and measure the dynamics of their subsystem entropies from the initial state $|+x\rangle^{\otimes N}$ or $|+y\rangle^{\otimes N}$ under a weak transverse field $B=2\pi\times 0.09\,$kHz. This time, in addition to the overall growth of the subsystem entropy, we further note that the initial entropy-growth rate is positively correlated with the spin-phonon coupling strength. Theoretically, if a spin pair is completely decoupled from the rest of the system, we expect its subsystem entropy to remain constant. In practice, we observe slower but nonzero increase in the subsystem entropy for ion pairs 2 and 4 near the nodes of the collective phonon mode ($b_i\approx 0$ in Eq.~(\ref{eq:Hamiltonian})). On the one hand, this is because the spin-phonon coupling strengths for these ions are not exactly zero, but only about $1/4$ of the stronger ones. On the other hand, there can also be contributions from coupling to other collective phonon modes, which have larger detunings (in this case we have detunings of $2\pi\times 3.6\,$kHz and $2\pi\times 4.3\,$kHz to the fourth and the second highest phonon modes) and thus smaller effects on the spin dynamics. Quantitatively, we choose more adjacent ion pairs and plot their fitted initial entropy-growth rates (using the first four time points as shown in Figs.~\ref{fig4}(a) and (b)) versus the coupling strengths $g$ in Figs.~\ref{fig4}(d) and (e). Here we define $g\equiv\sqrt{(g_1^2 + g_2^2)/2}$ as the root mean square of the spin-phonon couplings $g_1$ and $g_2$ of the two ions in a selected pair. Linear fits to these two plots, weighted by the inverse variance of each fitted entropy-growth rate, further give the correlation coefficients $r_x=0.85$ and $r_y=0.67$ and the corresponding p-values $p_x=1\times 10^{-3}$ and $p_y=9\times 10^{-3}$, indicating statistically significant positive correlation between the entropy-growth rate and the coupling strength. Finally, we extract the intercepts $-0.03\pm 0.04\,\mathrm{s}^{-1}$ and $0.06\pm 0.05\,\mathrm{s}^{-1}$ for these two plots, respectively. This suggests a relatively weak contribution of other decoherence sources to the initial entropy growth compared with the concerned spin-phonon coupling dynamics.

\section*{Discussion}
To sum up, in this work we demonstrate the quantum simulation of a Dicke-like model with nonuniform spin-boson coupling using a 2D trapped ion crystal. With the single-spin-resolved quantum state readout, we measure the global distribution of spin states which are generally challenging for direct classical simulation in the presence of nonuniformity, and we extract entropies of selected spin groups to characterize the relaxation dynamics of the system. In particular, we observe different entropy-growth rates in parameter regimes analogous to the normal phase and the superradiant phase of a homogeneous model, and for different ions in an engineered spin-boson coupling pattern.
Note that, although the complexity of the subensemble method used in this work to verify quantum simulation results is polynomial in $N$ for a fixed number of subensembles $M$, a reliable
approximation generally requires a large $M$ increasing with $N$ to avoid leakage outside the symmetric subspace of each subensemble during the evolution (see Methods for details). For a subsystem of $k=4$ ions we need $M=50$ subensembles for $N=200$, corresponding to a Hilbert-space dimension of more than 120
qubits, so that the overall dynamics remains out of reach of direct classical simulation.
Our quantum simulator allows for future studies of a two-stage relaxation dynamics within and outside the symmetric subspaces. It can also be combined with individually addressed quantum gates on 2D ion crystals \cite{hou2024individual} to study the thermalization dynamics of more complicated quantum states.

\begin{acknowledgements}
This work was supported by Quantum Science and Technology-National Science and Technology Major Project (Grant No. 2021ZD0301601, 2021ZD0301605), Beijing Science and Technology Planning Project (Grant No. Z25110100040000), the National Natural Science Foundation of China (Grant No. 12575021 and 12574541), Fundamental and Interdisciplinary Disciplines Breakthrough Plan of the Ministry of Education of China (JYB2025XDXM112), Tsinghua University Initiative Scientific Research Program, and the Ministry of Education of China. L.M.D. acknowledges in addition support from the New Cornerstone Science Foundation through the New Cornerstone Investigator Program. Y.K.W. acknowledges in addition support from the Dushi program from Tsinghua University. S.A.G. acknowledges support from the Shuimu Scholar program at Tsinghua University.
\end{acknowledgements}

\section*{Methods}

\subsection*{Detailed experimental setup and parameters}\label{Append:A}
\begin{figure}[tbp]
	\centering
	\includegraphics[width=0.8\linewidth]{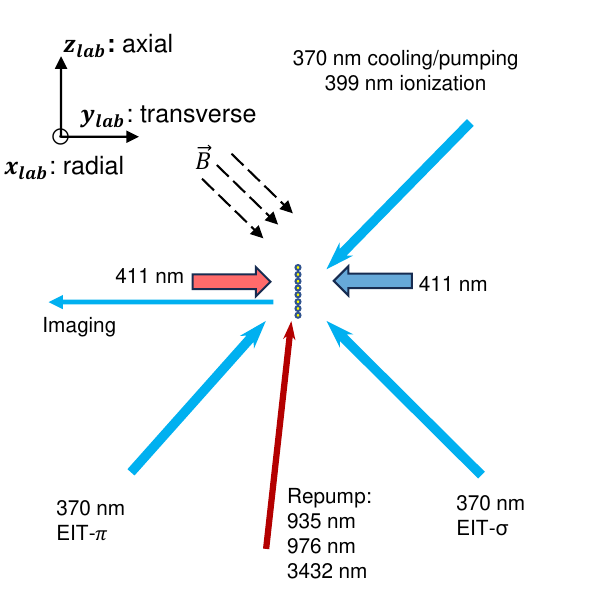}
	\caption {Configuration of the experimental setup. The 2D $^{171}\mathrm{Yb}^+$ crystal lies in the $x_{\mathrm{lab}}$-$z_{\mathrm{lab}}$ plane. The labeled beams are used as follows: 399 nm beam for ionization; one 370 nm beam for cooling, optical pumping, and fluorescence imaging; two other 370 nm EIT beams for cooling transverse phonon modes; two counter-propagating 411 nm beams for spin-phonon coupling and sideband cooling; 935, 976, and 3432 nm repump beams for the $D_{3/2}$, $D_{5/2}$, and $F_{7/2}$ manifolds, respectively. The magnetic field is oriented at $45^\circ$ to the crystal plane.}\label{fig:configuration}
\end{figure}
The detailed configuration of our experimental setup is shown in Fig.~\ref{fig:configuration}. The 2D ion crystal is on the $x_{\mathrm{lab}}$-$z_{\mathrm{lab}}$ plane of the trap, with the $z_{\mathrm{lab}}$ direction being the axial direction and the $x_{\mathrm{lab}}$ direction being one of the radial directions. A magnetic field of $4.5\,$Gauss is applied at $45^\circ$ to the crystal plane, and the polarization of the counter-propagating $411\,$nm laser beams stays in the plane spanned by the direction of magnetic field and the laser beams. This configuration is chosen to maximize the coupling strength between the $|S_{1/2}, F=0, m_F=0\rangle$ and the $|D_{5/2}, F=2, m_F=0\rangle$ states \cite{guo2024siteresolved}. A $370\,$nm laser beam for Doppler cooling, optical pumping and state detection is applied in the micromotion-free $y_{\mathrm{lab}}$-$z_{\mathrm{lab}}$ plane at angles to both axes. Two additional $370\,$nm laser beams, with a wave vector difference perpendicular to the 2D ion crystal, provide EIT cooling for the transverse phonon modes which are further cooled by the resolved sideband cooling using one of the two counter-propagating $411\,$nm laser beams. We further use $935\,$nm laser to repump the population in $D_{3/2}$ levels, and $976\,$nm and $3432\,$nm laser to repump the $D_{5/2}$ and $F_{7/2}$ levels.

For the $N=60$ and $N=200$ ion crystals, we set the trap frequencies at $(\omega_x,\omega_y,\omega_z) = 2\pi\times (0.54,1.96,0.10)\,$MHz. The average spin-phonon coupling strengths generated by the global $411\,$nm laser are set to $\bar{g}=2\pi\times 0.16\,$kHz for the $N=60$ crystal. For the $N=200$ crystal, we set $\bar{g}=2\pi\times 0.13\,$kHz for the experiments in Fig.~\ref{fig3}, and $\overline{|g|}=2\pi\times 0.25\,$kHz for Fig.~\ref{fig4}. (Note that for $g_i\equiv \eta\sqrt{N}b_i\Omega_i$ as we define, when coupled to phonon modes other than the COM mode, there can be positive and negative values.) The spatial nonuniformities are specified in the next section in Fig.~\ref{fig:nonuniform}. For the $N=10$ case, we modify the trap frequencies to $(\omega_x,\omega_y,\omega_z) =2\pi\times(0.62,1.92,0.13)\,$MHz to maintain a comparable distance between the ions, and the average spin-phonon coupling strength is set to $\bar{g}=2\pi\times 0.3\,$kHz.

\subsection*{Calibration of spatial nonuniformity}
\label{sec:nonuniform}
As described in the main text, the spin-boson coupling comes from a spin-dependent force as Eq.~(\ref{eq:Hamiltonian}) \cite{guo2024siteresolved}.

\begin{figure}[!tbp]
	\centering
	\includegraphics[width=\linewidth]{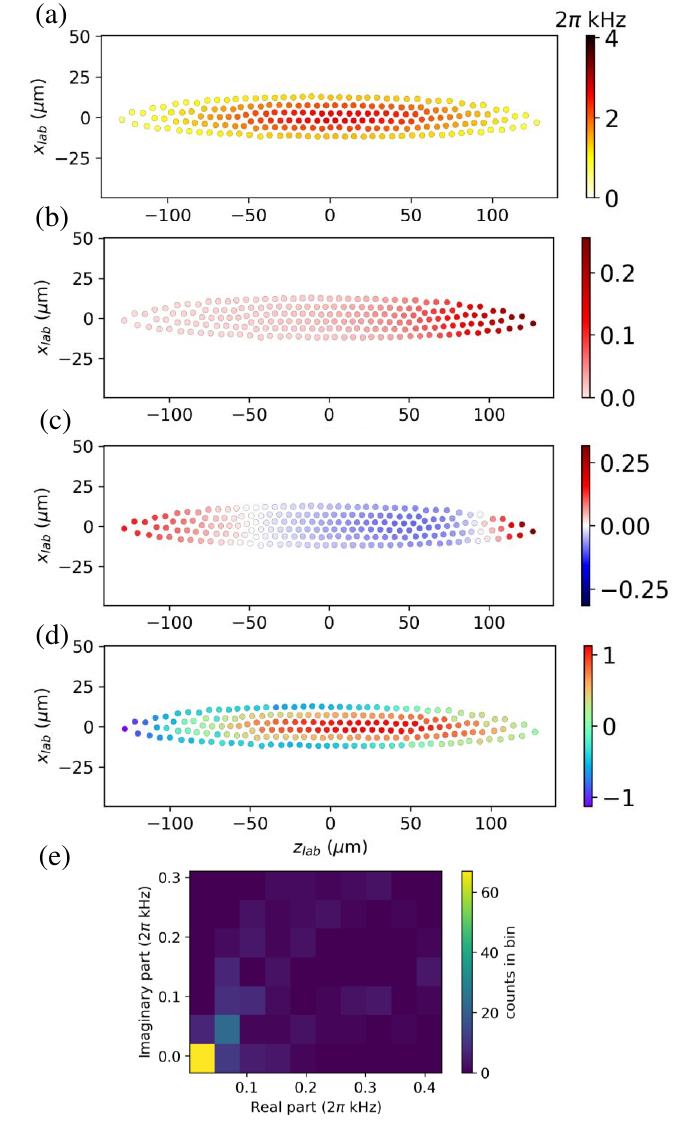}
	\caption{Calibrating spatial nonuniformity of spin-boson coupling for an $N=200$ ion crystal. (a) Spatial distribution of $411\,$nm-laser-induced light shift. (b) Mode structure of the COM mode. (c) Mode structure of the third highest mode. (d) Spatial distribution of motional phase. (e) 2D histogram of the complex coupling strength $g_i e^{i\varphi_i}$.} \label{fig:nonuniform}
\end{figure}

To compare the experimental results with theoretical calculation, we need to calibrate the spatial pattern of the spin-boson coupling. The light shift $\Omega_i$ can be obtained by measuring the resonant Rabi rates $\Omega_i^{(1)}$ and $\Omega_i^{(2)}$ of the optical transitions of individual $411\,$nm laser beams, fitting their spatial patterns by 2D Gaussian distributions to smooth out the experimental noise, and finally computing $\Omega_i = \Omega_i^{(1)} \Omega_i^{(2)}/(4\Delta)$ where $\Delta=2\pi\times 4\,$MHz is the $411\,$nm laser detuning when generating the spin-dependent force \cite{HamiltonianLearning2025}. The results are shown in Fig.~\ref{fig:nonuniform}(a).

To determine the phonon mode vector $b_i$ for a desired mode, we first fit the trapping potential up to the $4$th order anharmonic terms by the positions of individual ions and the frequencies of the collective phonon modes \cite{HamiltonianLearning2025}. Then we can theoretically calculate the desired mode vector as shown in Figs.~\ref{fig:nonuniform}(b) and (c) for the COM mode and the third highest mode.

Finally, for the site-dependent motional phase $\varphi_i$, we initialize all the ions in $|0\rangle$ and apply a $\pi/2$ pulse using one $411\,$nm laser, followed by another $\pi/2$ pulse with a random phase using the other laser. Then we compute the correlation between different ion pairs to get $(1/2)\cos(\varphi_i-\varphi_j)$ \cite{HamiltonianLearning2025}. By setting an arbitrary ion as the zero point, we obtain the distribution of the motional phase as shown in Fig.~\ref{fig:nonuniform}(d).

Combining all these nonuniform distributions together, we plot the 2D histogram of the complex coupling strength $g_i e^{i\varphi_i}$ in Fig.~\ref{fig:nonuniform}(e).
To quantify the nonuniformity, we characterize the complex coupling strength $g_i e^{i\varphi_i}$ by its standard deviation over the ion crystal: For the $N=200$ crystals used in Figs.~\ref{fig3} and \ref{fig4}, we obtain $2\pi\times 134\,$Hz and $2\pi\times 270\,$Hz, respectively, with mean values $2\pi\times 122\,e^{-0.56i}\,$Hz and $2\pi\times 121\,e^{2.29i}\,$Hz.

\subsection*{Numerical simulation by dividing into subensembles}\label{Append:C}
For a Dicke model with $N$ spins, the typical excitation number of the bosonic mode is $O(N)$, so the ideal homogeneous model has a dimension of $O(N^2)$, while a general nonuniform model requires a dimension of $O(N\cdot 2^N)$. It is thus not practical to solve the dynamics directly in the nonsymmetric Hilbert space for $N=60$ and $N=200$ cases using a classical computer.

\begin{figure}[htbp]
	\centering
	\includegraphics[width=\linewidth]{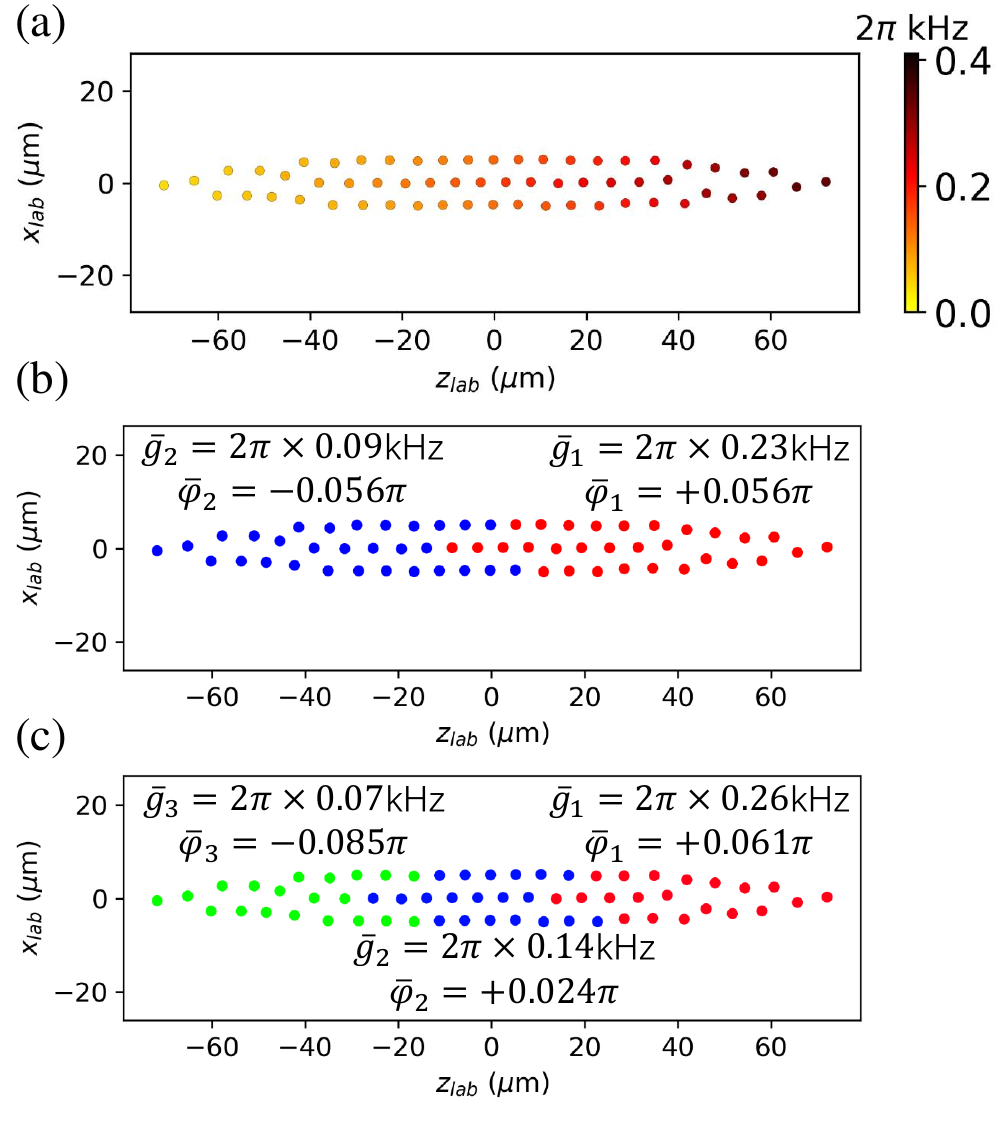}
	\caption{Division into different subensembles. (a) Spatial nonuniformity of the spin-boson coupling $g_i$ for an $N=60$ ion crystal. (b) Division into two subensembles based on smaller or larger values of $g_i$. (c) Division into three subensembles based on $g_i$.} \label{fig:nonuniform60ions}
\end{figure}

However, when coupled dominantly to the COM mode, the inhomogeneity mainly comes from the experimental imperfections which typically vary slowly in space. Then we can divide the spins into $M$ smaller subensembles, and expect nearly uniform coupling within each subensemble of $N_j=N/M$ spins. (Actually, as will be described later, even when coupling to other phonon modes with engineered spatial nonuniformity, we can still group the spins according to the closeness of their spin-phonon coupling strengths.) In this way, we can rewrite the Hamiltonian as
\begin{equation}
\begin{aligned}
H = &\frac{2}{\sqrt{N}} \sum_{j=1}^M g_j S_z^j(a e^{i\varphi_j}+a^\dag e^{-i\varphi_j}) \\
& + \frac{1}{\sqrt{N}}\sum_{j=1}^M  g_j N_j (a e^{i\varphi_j}+a^\dag e^{-i\varphi_j}) \\
& - \delta a^\dag a + 2B\sum_{j=1}^M S_x^j,
\end{aligned}\label{eq:subensemble}
\end{equation}
where we define components of total spin operators $S_{x(z)}^j\equiv \sum_{i\in I_j} \sigma_{x(z)}^i/2$ with $I_j$ being the index set for the $N_j$ spins in the subensemble $j$. We have also assumed $b_i=1/\sqrt{N}$ for the COM mode, and have absorbed its inhomogeneity into $g_j=\eta\Omega_j$ for each subensemble. Similarly, we assume a constant motional phase $\varphi_j$ for spins in the subensemble $j$.

In this way, we obtain a Hilbert space dimension of $O(N(N/M+1)^M)$, and achieve a polynomial scaling when fixing $M$ to be a small constant. In practice, given the values of $g_i$ and $\varphi_i$ for each spin, we can group them into subensembles according to the closeness of $g_i e^{i\varphi_i}$.
Specifically, for the numerical results in Figs.~\ref{fig1}(c) and (d), we divide the subensembles as shown in Fig.~\ref{fig:nonuniform60ions}(b) by sorting the magnitude of the coupling strengths in Fig.~\ref{fig:nonuniform60ions}(a) and taking the smaller and the larger halves. This gives us $\bar{g}_1 = 2\pi \times 0.23\,$kHz, $\bar{g}_2 = 2\pi \times 0.09\,$kHz, $\bar{\varphi}_1 = 0.056\pi$ and $\bar{\varphi}_2 = -0.056\pi$ for the two subensembles.
Note that these parameters are directly calculated from the experimentally calibrated values, and there is no additional fitting when plotting the numerical results. The standard deviations of the effective parameters within the two subensembles are $\sigma_{g_1}=2\pi\times0.06\,$kHz, $\sigma_{\varphi_1}=0.043\pi$ , $\sigma_{g_2}=2\pi\times0.03\,$kHz and $\sigma_{\varphi_2}=0.062\pi$, which quantify the quality of the subensemble approximation.

\begin{figure}[!tbp]
	\centering
	\includegraphics[width=\linewidth]{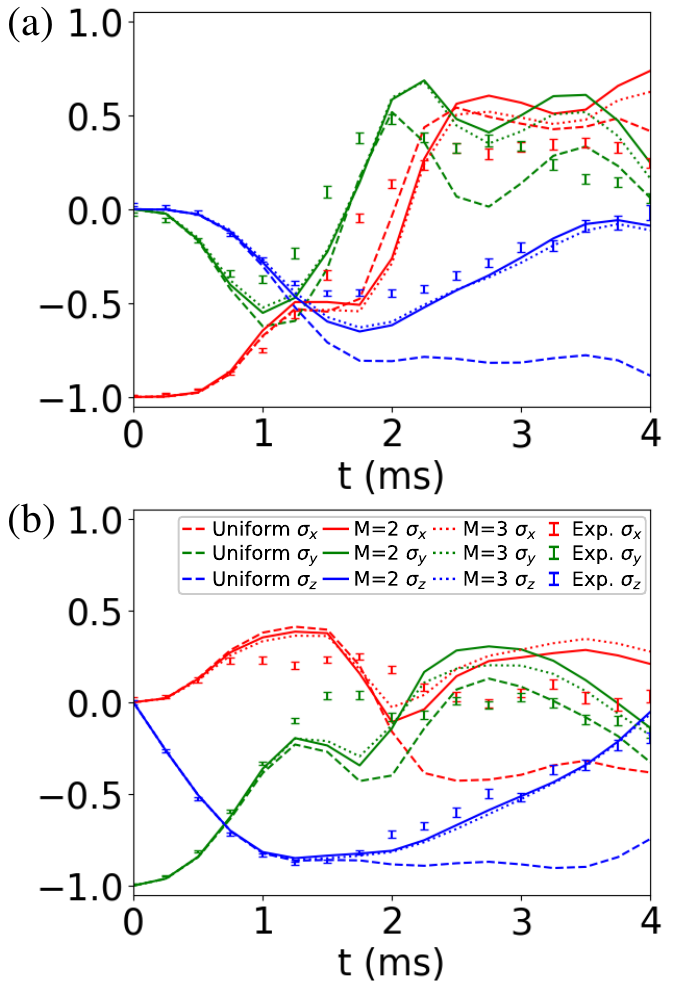}
	\caption{Convergence of the subensemble approximation. Measured average single-spin
observables (markers, the same as those in Figs.~1(c) and (d)) are compared with the numerical results using $M=1$ (homogeneous model), $M=2$ and $M=3$ subensembles, for the $N=60$ crystal initialized in (a)
$|-x\rangle^{\otimes N}$ and (b) $|-y\rangle^{\otimes N}$.}
	\label{fig:subensembleM3}
\end{figure}

By dividing into more and more subensembles, the approximation improves systematically and becomes exact in the limit $M=N$, although the computational cost then becomes intractable. As shown in Fig.~\ref{fig:nonuniform60ions}(c), we can similarly divide into three subensembles based on the magnitude of the coupling strengths, and obtain $\bar{g}_1 = 2\pi\times 0.26\,$kHz, $\bar{\varphi}_1 = +0.061\pi$, $\bar{g}_2 = 2\pi\times 0.14\,$kHz, $\bar{\varphi}_2 = +0.024\pi$, $\bar{g}_3 = 2\pi\times 0.07\,$kHz and $\bar{\varphi}_3 = -0.085\pi$, with suppressed standard deviations $\sigma_{g_1}=2\pi\times0.05\,$kHz, $\sigma_{\varphi_1}=0.040\pi$, $\sigma_{g_2}=2\pi\times0.03\,$kHz, $\sigma_{\varphi_2}=0.050\pi$, $\sigma_{g_3}=2\pi\times0.02\,$kHz and $\sigma_{\varphi_3}=0.051\pi$. In Fig.~\ref{fig:subensembleM3} we show similar plots as Figs.~\ref{fig1}(c) and (d) with additional theoretical results for $M=3$ subensembles as dotted curves.  As we can see, going from $M=1$ to $M=2$ improves the agreement between experimental and theoretical results substantially, whereas the change from $M=2$ to $M=3$ becomes small compared with the residual discrepancy, which is mainly caused by the drift of the experimental parameters. As a quantitative goodness-of-fit analysis, we compute the chi-squared statistic $\chi^2=\sum_{t,p}(O_{t,p}-E_{t,p})^2/\sigma_{t,p}^2$, where $O_{t,p}$ is the experimental value with standard
error $\sigma_{t,p}$ obtained from 200 experimental repetitions, $E_{t,p}$ is the theoretical value computed with $M=1$, $2$ and $3$ subensembles, $t$ runs over all the time points from $0$ to $4\,$ms in steps of $0.25\,$ms, and $p=x,y,z$ labels the average single-spin observables in the three Pauli bases. For the initial state $|-x\rangle^{\otimes N}$ we obtain $\chi^2_1=6.9\times10^3$, $\chi^2_2=3.7\times10^3$ and
$\chi^2_3=2.9\times10^3$ for $M=1,2,3$, respectively, and for $|-y\rangle^{\otimes N}$ we obtain $\chi^2_1=6.7\times10^3$, $\chi^2_2=2.5\times10^3$ and $\chi^2_3=2.0\times10^3$. Considering the compromise between the accuracy of the approximation and the numerical cost, in the main text we choose to present results up to $M=2$ subensembles.

\begin{figure}[!tbp]
	\centering
	\includegraphics[width=\linewidth]{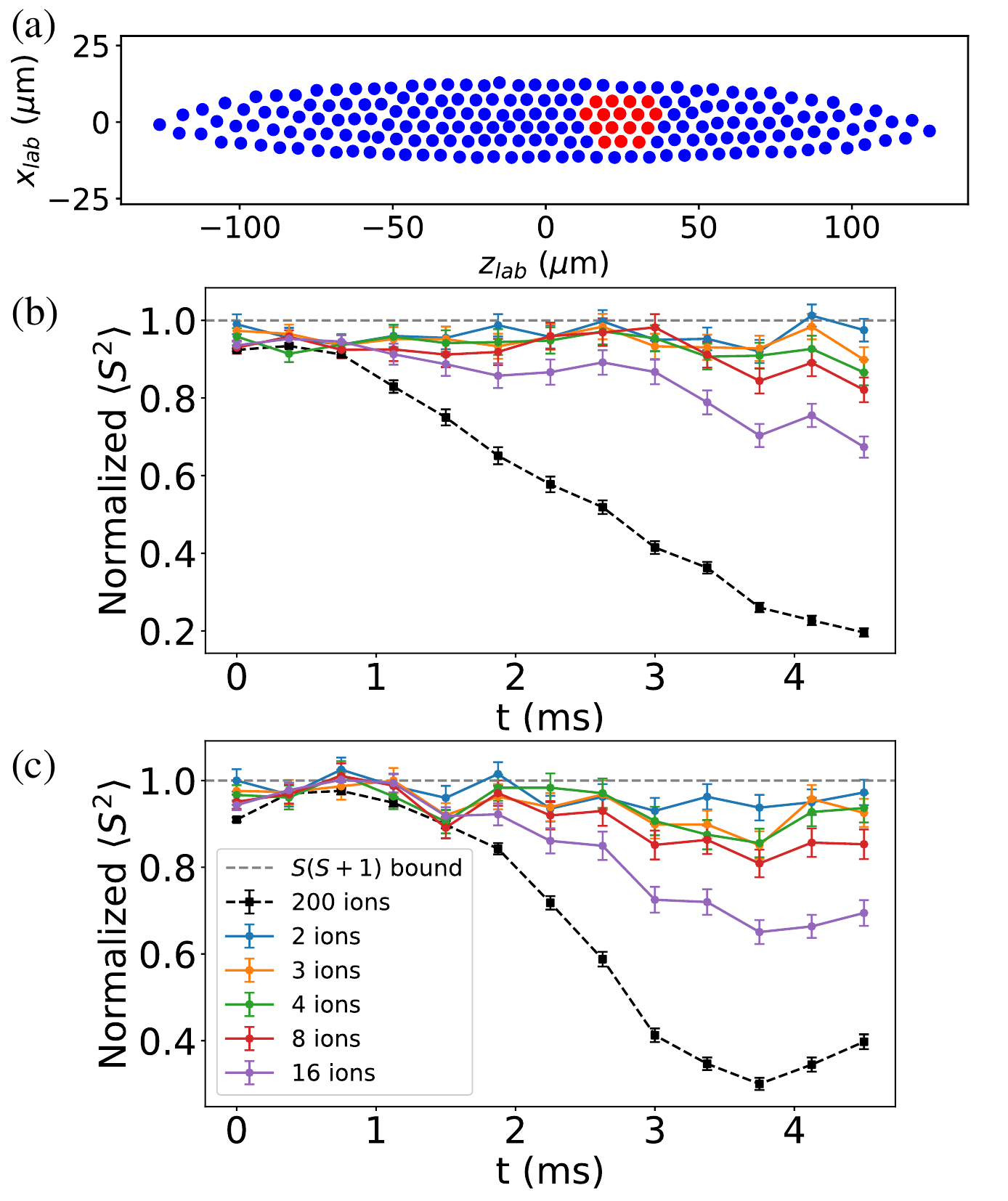}
	\caption{Evolution of total spin of subsystems. (a) Selection of $16$ spins in an $N=200$ crystal. (b) Evolution of total spin $\langle S^2\rangle$ (normalized by the maximal value $(k/2)(k/2+1)$) for subsystems with various sizes $k$ from the initial state $|+x\rangle^{\otimes N}$. (c) Similar plot for the initial state $|+y\rangle^{\otimes N}$.}
	\label{fig:totalS2}
\end{figure}

The accuracy of the local symmetric-subspace approximation can be further bounded by the leakage outside the subspace. For a group of $k$ spins, a leakage changes the total spin from $S^2=(k/2)(k/2+1)$ to at most $(k/2)(k/2-1)$ and at least zero, so the measured $\langle S^2\rangle$ provides upper and lower bounds on the leakage probability. From the measured dynamics of $\langle S^2\rangle$ as shown in Fig.~\ref{fig:totalS2} under the same experimental parameters as Fig.~\ref{fig3}, we
upper-bound the leakage probability (averaged over the initial states $|+x\rangle^{\otimes N}$ and $|+y\rangle^{\otimes N}$) to be $7.5\%$, $8.7\%$ and $13.9\%$ for the two-, three- and four-ion subsystems during the evolution, respectively. We note that decoherence such as spontaneous emission from the virtually excited $D_{5/2}$ levels to $F_{7/2}$ can also cause leakage. However, in this experiment we estimate only about $1\%$ population in $F_{7/2}$ per ion after $4\,$ms of evolution, which is subdominant.

\subsection*{Maximum likelihood estimation of spin states}
\label{Append:D}
To reconstruct the density matrices of a selected subgroup of spins, we perform single-shot measurement in random bases. In particular, we randomly and uniformly take $K$ points on the unit sphere (with $K=25$ for Fig.~\ref{fig3} and $K=16$ for Fig.~\ref{fig4}) and perform global rotations by microwave composite pulses to measure the spin states along these directions. For each direction, we take 50 samples of site-resolved spin states, and pick out the desired spins to reconstruct their reduced density matrices. Under the assumption of a symmetric subspace, we regard the measurement outcomes with the same number of spin excitations as the same Dicke state, and use the maximum likelihood method to find the most probable quantum state to give the measured outcomes \cite{PhysRevA.63.040303}. To completely reconstruct the $k$-spin density matrices, we further distinguish all the site-resolved measurement results and again perform the maximum likelihood estimation.

The error bars of the reconstructed subsystem entropies in Figs.~\ref{fig3} and \ref{fig4} are estimated by bootstrap resampling. For each time point we resample the experimental shots with replacement, repeat the density-matrix reconstruction and the entropy evaluation for 20 times, and take the standard deviation as the error bar. This estimation naturally covers the finite-sample effect, but may not be able to capture the bias of the maximum likelihood estimation itself. The latter is known to underestimate the purity for states close to a pure state \cite{PhysRevA.95.022107}, which explains the small residual entropy we obtain for the initial product state. We do not apply a state-preparation-and-measurement (SPAM) error correction as the electron shelving method already gives a SPAM error below $1\%$ per ion, which is not significant for subsystems of up to four spins. The uncertainties of other fitting results are also obtained by the standard deviation over 20 rounds of bootstrap resampling.

\bibliography{reference_npj}

\section*{Data Availability}
The data that support the findings of this article are openly available at {\rmfamily \url{https://doi.org/10.6084/m9.figshare.31307455}}.

\section*{Code Availability}
The codes are available from the corresponding authors upon reasonable request.

\section*{Competing interests}
C.Z. is affiliated with HYQ Co. B.X.Q., L.H., Z.C.Z., Y.K.W. and L.M.D. hold shares with HYQ Co. The other authors declare no competing interests.

\section*{Author contributions}
S.A.G., J.Y., J.Y.T., L.Z., Y.Y.C., Y.L.X., C.Z., Y.J., B.X.Q., L.H. and Z.C.Z. carried out the experiment. L.M.D. and Y.K.W. proposed and supervised the project. S.A.G., J.Y., Z.W.Z. and Y.K.W. analyzed the data and did the associated theory. Y.K.W., S.A.G., J.Y. and L.M.D. wrote the manuscript.

\end{document}